\documentclass{PoS}

\usepackage{graphicx}
\usepackage{subfigure}

\def\rcite#1{Ref.~\cite{#1}}

\def\eqn#1{\label{eq:#1}}

\def\eq#1{Eq.~(\ref{eq:#1})}

\def\figref#1{Fig.~\ref{fig:#1}}
\def\Figref#1{Figure~\ref{fig:#1}}

\def\tabref#1{Table~\ref{tab:#1}}

\title{Gradient Flow Analysis on MILC HISQ Ensembles}

\ShortTitle{Gradient Flow Analysis on MILC HISQ Ensembles}

\author{A.~Bazavov$^a$\thanks{Present address:~{Department of Physics and Astronomy, University of Iowa, Iowa City, IA, USA}}, C.~Bernard$^b$, \speaker{N.~Brown}$^b$, C.~DeTar$^c$, J.~Foley$^c$, Steven~Gottlieb$^d$, U.M.~Heller$^e$, J.E.~Hetrick$^f$, J.~Komijani$^b$, J.~Laiho$^g$, L.~Levkova$^c$, M.~Oktay$^c$, R.L.~Sugar$^h$, D.~Toussaint$^i$, R.S.~Van~de~Water$^j$, R.~Zhou$^j$ \,(MILC Collaboration)\\

$^a$ Physics Department, Brookhaven National Laboratory, Upton, NY 11973, USA\\
$^b$ Department of Physics, Washington University, St. Louis, MO 63130, USA\\
$^c$ Department of Physics and Astronomy, University of Utah, Salt Lake City, UT 84112, USA\\
$^d$ Department of Physics, Indiana University, Bloomington, IN 47405, USA\\
$^e$ American Physical Society, One Research Road, Ridge, NY 11961, USA\\
$^f$ Physics Department, University of the Pacific, Stockton, CA 95211, USA\\
$^g$ Department of Physics, Syracuse University, Syracuse, NY 13244, USA\\
$^h$ Physics Department, University of California, Santa Barbara, CA 93106, USA\\
$^i$ Physics Department, University of Arizona Tucson, AZ 85721, USA\\
$^j$ Fermi National Accelerator Laboratory, Batavia, IL 60510, USA\\
E-mail: \email{brownnathan@wustl.edu}
}

\abstract{We report on a preliminary scale determination with gradient-flow techniques on the $N_f=2+1+1$ HISQ ensembles generated by the MILC collaboration. The ensembles include four lattice spacings, ranging from 0.15 to 0.06 fm, and both  physical and unphysical values of the quark masses. The scales $\sqrt{t_0}/a$ and $w_0/a$ are computed using Symanzik flow and the cloverleaf definition of $\langle E \rangle$ on each ensemble. Then both scales and the meson masses $aM_\pi$ and $aM_K$ are adjusted for mistunings in the charm mass. Using a combination of continuum chiral perturbation theory and a Taylor series ansatz in the lattice spacing, the results are simultaneously extrapolated to the continuum and interpolated to physical quark masses. Our preliminary results are $\sqrt{t_0}=0.1422(7)$ fm and $w_0=0.1732(10)$ fm. We also find the continuum mass-dependence of $w_0$.}

\FullConference{The 32nd International Symposium on Lattice Field Theory,\\
		23-28 June, 2014\\
		Columbia University New York, NY}

\begin{document}

\vspace{-1mm}
\section{Introduction}
\vspace{-2mm}
Scale setting holds central importance in lattice QCD for two reasons. First, the continuum extrapolation of any quantity requires precise determination of the relative scale between ensembles with different bare couplings. Second, the precision of any dimensionful quantity in physical units is limited by the precision of the scale in physical units (the {\it absolute scale}).

Any dimensionful quantity that is finite in the continuum limit may be used for scale-setting. The relative scale may be set by calculating a dimensionful quantity and comparing its value in lattice units at different lattice spacings for the same physical quark masses. For absolute scale setting, one needs to compare the quantity in lattice units to its value in physical units, which may either be determined directly from experiment, or indirectly, by comparison to an experimentally-determined quantity.  A non-experimental quantity may be useful for scale setting if it can be determined on the lattice with small statistical and systematic errors for relatively small computational cost. This may give a large gain in control over continuum extrapolations, at the price, perhaps, of a small additional error from the extra step of comparison to an experimentally-determined quantity.  This has led to the consideration of theoretically-motivated, but not experimentally measurable, quantities such as $r_0$ \cite{r0} and $r_1$ \cite{r1}, $F_{p4s}$ \cite{fp4s}, and, more recently, $\sqrt{t_0}$ \cite{luscher_t0} and $w_0$ \cite{bmw} from gradient flow.
Here we describe our calculation of gradient-flow scales on the MILC (2+1+1)-flavor HISQ ensembles \cite{hisq1,hisq2}.

\vspace{-1mm}
\section{Theoretical Details}
\vspace{-2mm}
Gradient flow \cite{smoothing_origin, luscher_origin} is a smoothing of the original gauge fields $A$ towards stationary points of the action. The new, smoothed gauge fields $B(t)$ are functions of the `flow time' $t$ and updated according to the diffusion-like equation,
	\[ \frac{d B_\mu}{dt}  = \partial_\nu G_{\nu\mu} + \left[B_\nu,\, G_{\nu\mu} \right]\,, \hspace{3mm} 
	   G_{\nu\mu} = \partial_\nu B_\mu - \partial_\mu B_\nu + \left[ B_\nu,\, B_\mu \right]\,, \hspace{3mm} 
	   B_\mu(0) = A_\mu\,. \]

The process of gradient flow introduces a dimensionful independent variable, the flow time. One may define a scale by choosing a reference time at which a chosen dimensionless quantity reaches a predefined value. If the dimensionless quantity is also finite in the continuum limit, then the reference time scale $t_0$ will be independent of the lattice spacing up to discretization corrections. One of the easiest, dimensionless quantities to calculate with only gauge fields is the product of the energy density and squared flow time $t^2 \langle E(t) \rangle$. The energy density is proven to be finite to next-to-leading order (when expressed in terms of renormalized quantities) \cite{luscher_t0}, so $t^2 \langle E(t) \rangle$ is a suitable candidate for setting the scale. A fiducial point $c$ is chosen, and the reference scale is defined by $t_0^2 \langle E(t_0) \rangle = c$. The fiducial point is chosen so that for simulated lattice spacings and volumes, the reference timescale $t_0$ is large enough to reduce discretization effects but small enough to avoid finite volume corrections.  The value of $c=0.3$ was found empirically to satisfy these requirements \cite{luscher_t0, bmw}. Further empirical evidence suggests that discretization effects have little impact on the slope of  $t^2 \langle E(t) \rangle$ at larger flow times \cite{bmw}. Assuming the property is general, one may define a scale-setting quantity $w_0$ with smaller discretization effects: $t \frac{d}{dt} t^2 \langle E(t) \rangle = c$ with $t$ evaluated at $w_0^2$. Again, the value of the fiducial point $c=0.3$ is chosen to avoid discretization and finite volume effects.

Because both scales $t_0$ and $w_0$ are defined in terms of the energy density $\langle E(t) \rangle$, and the energy density is a local, gauge-invariant quantity, chiral perturbation theory can be applied to determine the quark mass dependence. The expansion for $\sqrt{t_0}$ in the $N_f=2+1$ case \cite{chipt} is 
\begin{eqnarray}
	\sqrt{t_0} & = & \sqrt{t_{0,ch}}\;\Bigg\{1 + c_1 \frac{2M_K^2 + M_\pi^2}{(4\pi f)^2} + \frac{1}{(4\pi f)^4} \left( (3c_2-c_1)M_\pi^2\mu_\pi + \phantom{\frac{.}{.}}4c_2M_K^2\mu_K \right. \eqn{t0-chpt} \\
		& + &  \left. \frac{1}{3}c_1 (M_\pi^2-4M_K^2)\mu_\eta + c_2M_\eta^2\mu_\eta + c_4(2M_K^2+M_\pi^2)^2 + c_5(M_K^2-M_\pi^2)^2 \right) \Bigg\}\ , \nonumber
\end{eqnarray}
\noindent
where $ t_{0,ch}$ is the value of $t_0$ in the chiral limit, the chiral logarithms are represented with the shorthand $\mu_Q = M_Q^2 \log{(M_Q/\mu)^2}$, and $c_i$ are low energy constants (LECs) that depend on the flow time. 
The scale $w_0$ has the same expansion form, but with different coefficients $c_i$.

One can generalize \eq{t0-chpt} to staggered chiral perturbation theory in order to explicitly take into account discretization effects from staggered taste-symmetry violations.  In this paper, however, we have used simple polynomial expansions in lattice-spacing effects for two reasons. First, as will be evident below, the quark-mass dependence of the gradient flow scales is small, and nontrivial staggered effects would appear only in the chiral logarithms, namely at NNLO. For HISQ quarks, such effects are very small. Second, the number of undetermined coefficients in staggered chiral perturbation theory expansions would be large compared to the number of independent data points available for interpolations. Unlike analyses of pseudoscalar masses or decay constants, here we have no valence quarks whose masses could be varied to increase the dataset.

Mistunings of the charm quark mass on our ensembles vary between 1\% and 11\%. It is therefore important to account for the leading order corrections in the charm mass to the quantities we consider. Given any low-energy quantity $Q$ that is proportional to a power $p$ of $\Lambda_{\rm QCD}^{(3)}$ in the effective three-flavor low-energy theory,  the leading order heavy-quark mass dependence can be determined using the relation between $\Lambda_{\rm QCD}^{(3)}$  and $\Lambda_{\rm QCD}^{(4)}$ of the four flavor theory \cite{charm1,charm2}.  
The derivative of $Q$ with respect to $m_c$ at leading order is proportional to $pQ/m_c$. For a dimensionless ratio, such as $F_{p4s}/w_0^{-1}$, where both dimensional quantities share the same power $p$, the leading-order dependence on $m_c$ will cancel. 
However, for dimensionless ratios where the powers are not identical, such as $M_\pi/F_{p4s}$ where $M_\pi\propto (m_l\Lambda_{\rm QCD}^{(3)})^{(1/2)}$, 
dependence on $m_c$ will remain. 


\vspace{-1mm}
\section{Analysis}
\vspace{-2mm}
\label{sec:analysis}

\tabref{gfAll} shows the results for $\sqrt{t_0}/a$ and $w_0/a$ on the HISQ ensembles\cite{hisq1,hisq2}. For the ensembles with the smallest lattice volumes, all configurations are included in the computation. As the volumes and cost become larger, a fraction of the configurations are analyzed. These subsets have uniform spacing between configurations. The total number of generated configurations, number of configurations in the gradient flow calculation, and molecular dynamics time separation between the included configurations are tabulated for each ensemble in \tabref{gfAll}. The error listed with each scale is statistical and determined by a jackknife analysis. The jackknife bin size is set to be at least twice the integrated autocorrelation length of the energy density, conservatively determined in previous work \cite{prelim}.

\begin{table}
\small
\begin{center}
\subfigure[Physical strange mass ensembles ($m'_s=m_s$)]{%
\begin{tabular}{|l|l|c|c|c|c|}
\hline
$\approx a$(fm) & $m_l'/m_s'$ & $N_{sim}/N_{gen}$ & $\tau$ & $\sqrt{t_0}/a$ & $w_0/a$ \\
\hline
$0.15$ & $1/5$  & $1020/1020$ & $5$ & $1.1004(05)$ & $1.1221(08)$ \\
$0.15$ & $1/10$ & $1000/1000$ & $5$ & $1.1092(03)$ & $1.1381(05)$ \\
$0.15$ & $1/27$ & $999/1000$  & $5$ & $1.1136(02)$ & $1.1468(04)$ \\
\hline
$0.12$ & $1/5$            & $1040/1040$ & $5$   & $1.3124(06)$ & $1.3835(10)$ \\
$0.12$ & $1/10\ (32^3\times64)$ & $999/1000$  & $5$   & $1.3228(04)$ & $1.4047(09)$ \\
$0.12$ & $1/10\ (40^3\times64)$ & $1000/1028$ & $5$   & $1.3226(03)$ & $1.4041(06)$ \\
$0.12$ & $1/27$           & $34/999$    & $140$ & $1.3285(05)$ & $1.4168(10)$ \\
\hline
$0.09$ & $1/5$  & $102/1011$ & $50,60$ & $1.7227(08)$ & $1.8957(15)$ \\
$0.09$ & $1/10$ & $119/1000$ & $36$    & $1.7376(05)$ & $1.9299(12)$ \\
$0.09$ & $1/27$ & $67/1031$  & $32,48$ & $1.7435(05)$ & $1.9470(13)$ \\
\hline
$0.06$ & $1/5$  & $127/1016$ & $48$ & $2.5314(13)$ & $2.8956(33)$ \\
$0.06$ & $1/10$ & $38/1166$  & $96$ & $2.5510(14)$ & $2.9478(31)$ \\
$0.06$ & $1/27$ & $49/583$   & $48$ & $2.5833(07)$ & $3.0119(19)$ \\
\hline
\end{tabular}
\label{tab:gfPhys}
}%

\subfigure[Non-physical strange mass ensembles ($m_s'\not=m_s$)]{%
\begin{tabular}{|l|l|c|c|c|c|}
\hline
$ m_l'/m_s$ & $m_s'/m_s$ & $N_{sim}/N_{gen}$ & $\tau$ & $\sqrt{t_0}/a$ & $w_0/a$ \\
\hline
$0.10$  & $0.10$ & $102/1020$ & $20$ & $1.3596(06)$ & $1.4833(13)$ \\
$0.10$  & $0.25$ & $204/1020$ & $20$ & $1.3528(04)$ & $1.4676(10)$ \\
$0.10$	& $0.45$ & $205/1020$ & $20$ & $1.3438(05)$ & $1.4470(10)$ \\
$0.10$	& $0.60$ & $107/1020$ & $20$ & $1.3384(08)$ & $1.4351(16)$ \\
\hline
$0.175$ & $0.45$ & $133/1020$ & $20$ & $1.3385(05)$ & $1.4349(13)$ \\
$0.20$  & $0.60$ & $255/1020$ & $20$ & $1.3297(06)$ & $1.4170(12)$ \\
$0.25$  & $0.25$ & $255/1020$ & $20$ & $1.3374(07)$ & $1.4336(14)$ \\
\hline
\end{tabular}
\label{tab:gfUnphys}
}%
\end{center}
\vspace{-5mm}
\caption{The gradient flow scales $\sqrt{t_0}/a$ and $w_0/a$ on the HISQ ensembles. For the physical strange mass ensembles, the first two columns list the approximate lattice spacing and ratio of light to strange sea-quark mass with the lattice dimensions appended as needed to uniquely identify each ensemble. For the non-physical strange mass ensembles, the first two columns list the ratio of light or strange sea quark mass to the physical strange quark mass. The lattice spacing is $\approx 0.12$fm for these ensembles. For all ensembles, the next column shows the ratio of number of configurations included in the gradient flow calculation to the number of configurations in the ensemble. The fourth column lists the molecular dynamics time separation $\tau$ between configurations included in the gradient flow calculation. Multiple values are listed for cases where independent streams of the same ensemble did not have the same $\tau$.}
\label{tab:gfAll}
\vspace{-3mm}
\end{table}


Using all of the ensembles listed in \tabref{gfAll}, we perform a combined continuum extrapolation and interpolation to physical quark mass of $\sqrt{t_0}F_{p4s}$ and $w_0F_{p4s}$. $F_{p4s}$ is the pseudoscalar decay constant with degenerate valence quarks $m_{\rm v}=0.4m_s$ and physical sea-quark masses \cite{fp4s}. To perform the extrapolation, there are three functional forms that must be chosen: quark mass terms, lattice spacing terms, and terms that combine both (cross-terms). For the mass dependence we use the chiral expansion outlined in \eq{t0-chpt} with $M_\pi^2$ and $M_K^2$ as independent variables standing in for the quark mass dependence. We consider fits with the expansion up to LO, NLO, or NNLO. The HISQ action has leading discretization errors of ${\cal O}(\alpha_s a^2)$.  However ${\cal O}(a^2)$ tree-level discretization errors are introduced by the action used for the flow and by the observable $\langle E\rangle $ \cite{zeugen}. For the lattice-spacing dependence in our fits, we therefore use a Taylor series ansatz in both $a^2$ and $\alpha_sa^2$. Every fit includes at least $a^2$ with optional additional powers up to $a^6$. Terms with $\alpha_sa^2$ are optionally included up to the same order as $a^2$. We consider only products of chiral and lattice spacing terms whose total order is no higher than the largest non cross-term included in the fit function. For counting orders, we consider $(\Lambda_{\rm QCD}a)^2 \sim (M/(4\pi f))^2$. 

We also consider two restrictions of the dataset. We try fits that drop the coarsest, $a\approx 0.15$ fm, ensembles, and then consider lattice spacing dependence only up to order $a^4$, as the three remaining lattice spacings would otherwise be parameterized by four variables. The kaon mass imposes the second restriction. The lighter-than-physical strange-quark ensembles cover strange masses all the way down to $0.1m_s$. So, we consider seven different lower bounds for the kaon mass, ranging from including all ensembles to including only ensembles with physical strange-quark mass.

Overall, there are three chiral expansions, nine discretization expansions with the $a\approx 0.15$ fm lattices, five discretization expansions without the $a\approx 0.15$ fm lattices, and seven choices of lower bound for the kaon mass. This produces a total of $3\times(9+5)\times7=294$ different fits. We use the $p$-value and proximity to the finest ($a\approx0.06$ fm) physical-quark-mass ensemble to gauge the fits' acceptability. For both $\sqrt{t_0}$ and $w_0$, the fits with p-values$>0.01$ also pass close to the point from the $a\approx0.06$ fm physical ensemble (deviations of less than $2\sigma_{\rm stat}$). For all further analysis, we only analyze these `acceptable' fits, of which there are 70 for $\sqrt{t_0}$ and 45 for $w_0$.

\begin{figure}
\begin{center}
	\subfigure{%
		\includegraphics[width=0.5\textwidth]{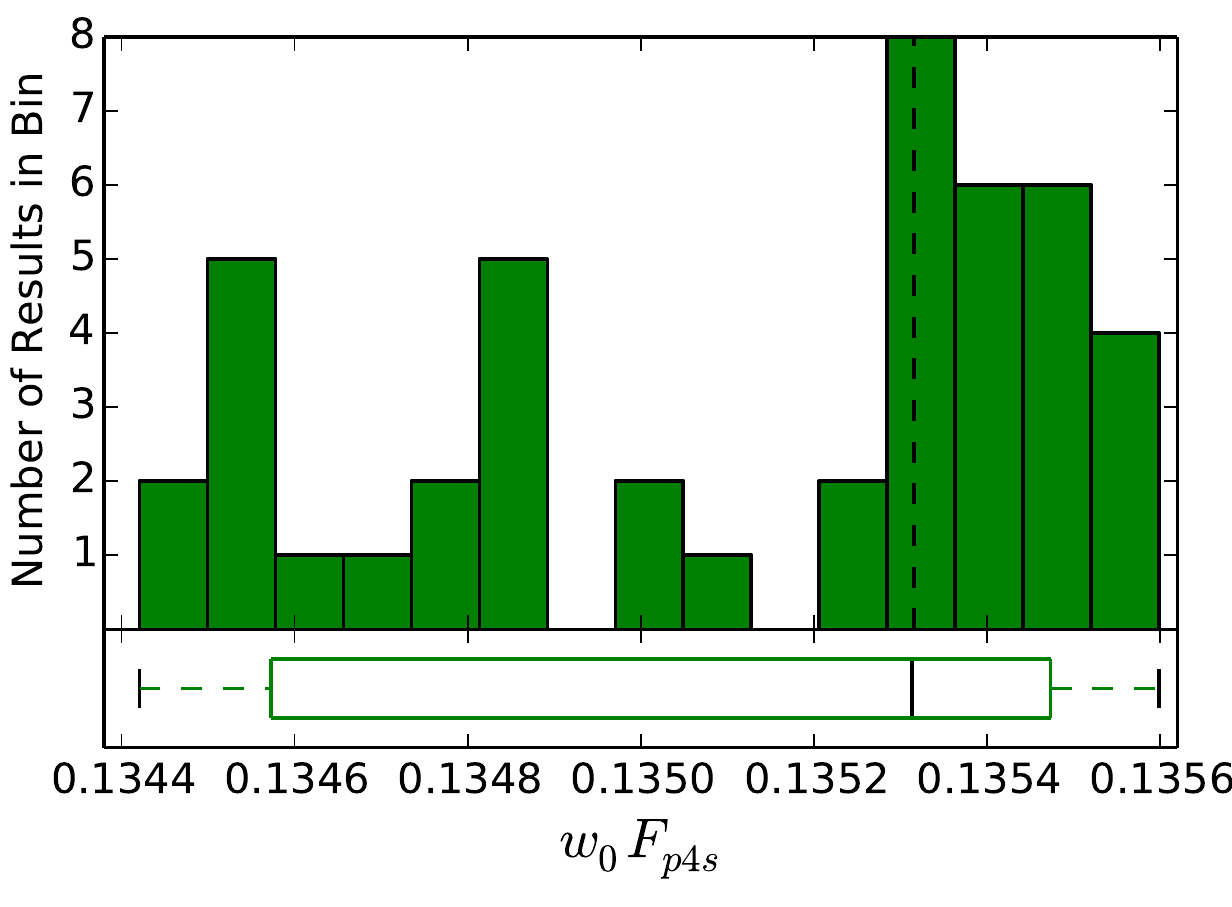}
	}%
	\subfigure{%
		\includegraphics[width=0.5\textwidth]{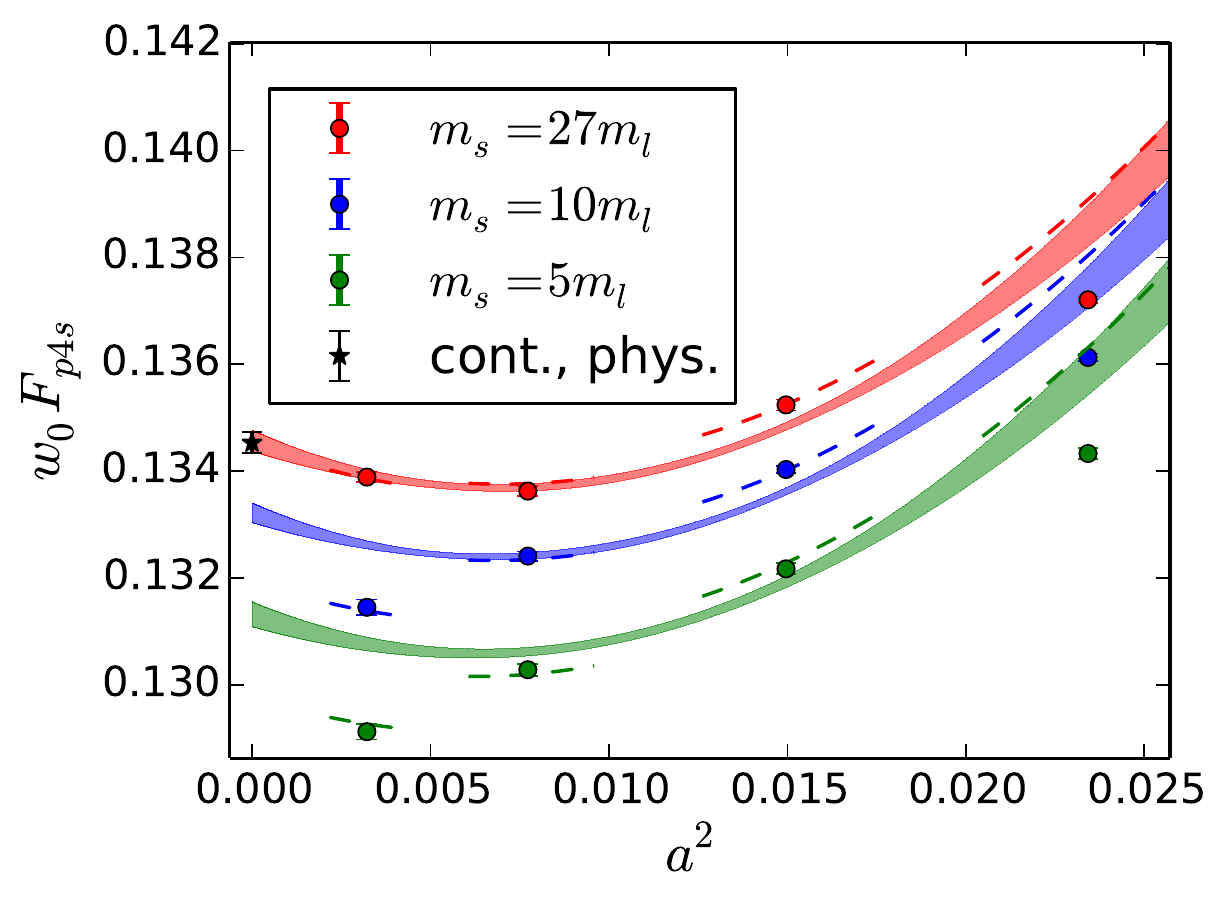}
	}
\end{center}
\vspace{-5mm}
\caption{Left) Histograms of the continuum extrapolations for $w_0 F_{p4s}$ for all fits with $p>0.01$. The box and error bars along the bottom denote the minimum, median, maximum, and central 68\% of the distribution. The vertical dashed line marks the continuum result for the central fit chosen from this subset.
Right) The central fit to $w_0F_{p4s}$ plotted as a function of $a^2$. The continuum value for $w_0$ at physical quark masses is indicated by the black star. Only $m'_s=m_s$ ensembles are plotted, but the fits include all $m'_s \le m_s$ ensembles. Dashed lines represent the fit at the quark mass value and lattice spacing of each ensemble, while the solid lines show the fit as a function of lattice spacing after retuning to the physical strange-quark mass and the desired ratio of $m'_l/m_s$ specified in the legend.}
\label{fig:histextrap}
\vspace{-3mm}
\end{figure}

We take all the acceptable fits and create a histogram of the continuum extrapolated values. The histogram for $w_0F_{p4s}$ is illustrated in \figref{histextrap}. The histograms for both $\sqrt{t_0}F_{p4s}$ and $w_0F_{p4s}$ show a bimodal distribution roughly distinguished by the presence or absence of $\alpha_sa^2$ terms. For $\sqrt{t_0}F_{p4s}$ the median and mean fall in the range where $\alpha_s$ is not used, while the opposite is true for $w_0F_{p4s}$. At this time neither mode can be eliminated, so we use the full width of the entire histogram as an estimate of the systematic error from the choice of fit.

A central fit is chosen to be a fit close to the median of the distribution, with $p>0.1$, and at least twice as many data points as parameters. For $\sqrt{t_0}$ the central fit contains no $\alpha_s$ terms but does contain $a^2$, $a^4$, and up to NNLO chiral terms. The fit has $\chi^2/{\rm dof} = 12.9/9$, $p=0.167$, and passes $0.7\sigma$ below the physical-mass 0.06 fm ensemble. For $w_0$ the central fit contains $\alpha_sa^2$, $a^2$, and up to NNLO chiral terms. The fit has $\chi^2/{\rm dof} = 12.5/10$, $p=0.255$, and passes $0.6\sigma$ below the physical-mass 0.06 fm ensemble. For both $\sqrt{t_0}$ and $w_0$ the $a\approx0.15$ fm ensembles were dropped, but all values of the kaon mass were included. A plot of the central fit for $w_0F_{p4s}$ is shown in \figref{histextrap}, where $\alpha_s$ is interpolated as a function of $a^2$ to make the plot. The accuracy of the fit can be judged using the dashed lines, which represent the fit evaluated at the same masses and lattice spacing as the data points. The three solid bands plotted across the entire figure show the lattice spacing dependence at fixed, re-tuned quark masses. 

\begin{figure}
\begin{center}
\includegraphics[width=0.8\textwidth]{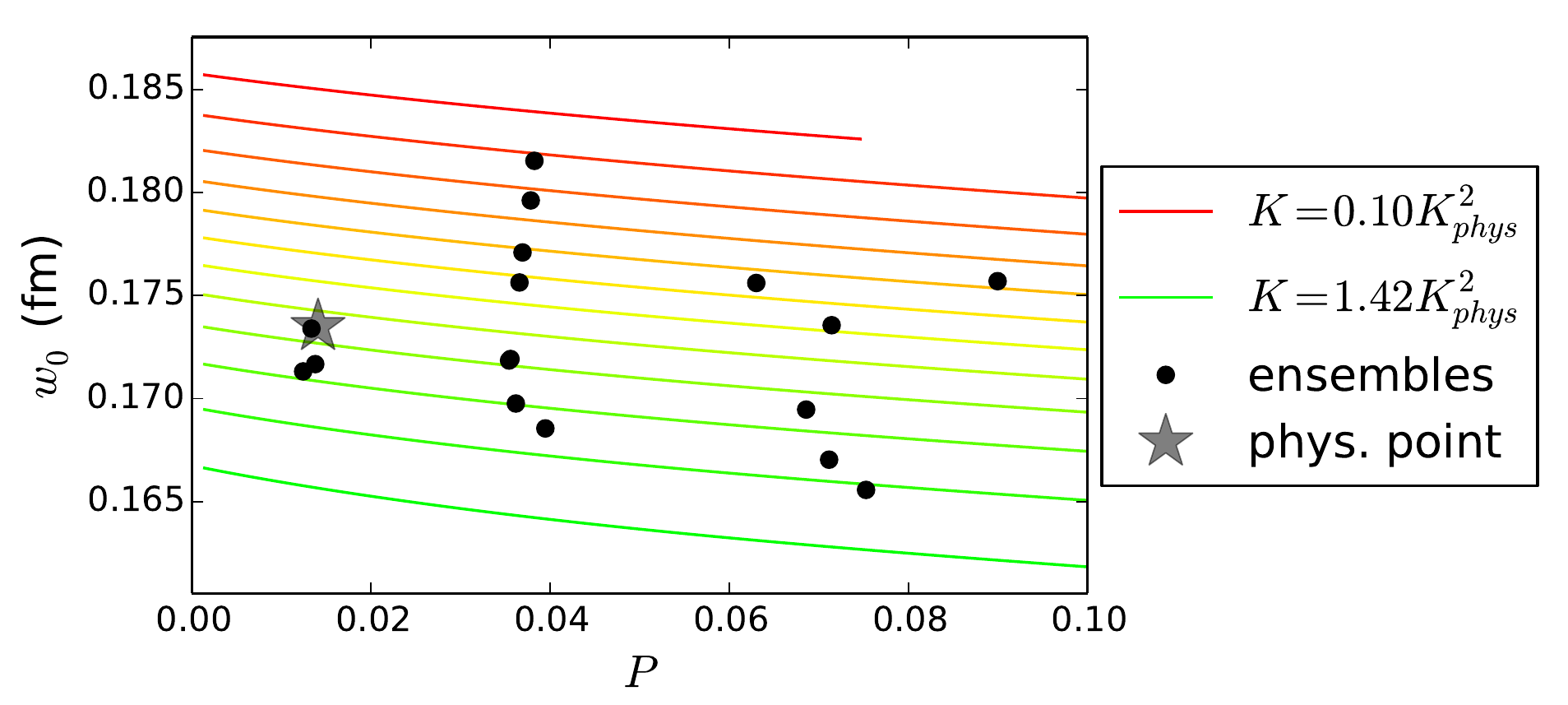}
\end{center}
\vspace{-5mm}
\caption{The continuum mass dependence of $w_0$ as a function of the pion mass $P=(w_0M_\pi)^2$ for fixed values of the kaon mass $K=(w_0M_K)^2$. The black circles illustrate $(w_0F_{p4s})/F_{p4s}^{\rm phys}$ for all ensembles $a<0.15fm$. The black star indicates the value of $w_0$ at the physical pion and kaon masses.
\vspace{-3mm}
\label{fig:mass}}
\end{figure}

Using the central fit and data set, we determine the continuum mass dependence of $w_0$. This function is useful for explicit comparisons of mass dependence to other scale setting quantities and for predicting the scales on future ensembles. 
The functional form of the mass dependence $f(P,K)$ is chosen to be the same as the chiral perturbation theory expansion to NNLO, in agreement with the  central fit. The coefficients are determined by solving the implicit equation $w_0=f(P=(w_0M_\pi)^2,K=(w_0M_K)^2)$ numerically for $w_0$, where the function $f$ is constructed by scaling the optimal fit by $F_{p4s}$(fm) and evaluating in the continuum. 
\Figref{mass} plots this function over a large range of values for the meson masses $P$ and $K$. Values corresponding to the HISQ ensembles and physical mass point are overlaid to give a sense of the range of data that went into constructing the plot.

\vspace{-1mm}
\section{Results and Conclusion}
\vspace{-2mm}
By evaluating $\sqrt{t_0}F_{p4s}$ and $w_0F_{p4s}$ at physical quark mass, extrapolating to the continuum,
and dividing by the physical value of $F_{p4s}=153.90(9)({}^{+21}_{-28})$ MeV determined in \rcite{fp4s}, we compute our preliminary estimate of the gradient flow scales in physical units.
The central values are taken from the center of the histograms. The first error in each case is statistical, and the remaining errors are systematic from: the choice of fit form to $\sqrt{t_0}F_{p4s}$ or $w_0F_{p4s}$ (which includes continuum extrapolation and mass-interpolation errors), continuum-extrapolation and mass-interpolation errors in $F_{p4s}$, residual finite volume effects on $F_{p4s}$ (through $f_\pi$), and the error in $F_{p4s}$ coming from the experimental error in $f_\pi$ \cite{PDG}. 
\begin{eqnarray*}
\sqrt{t_0} &=& 0.1422 (2)_{stat} (5)_{extrap} ({}^{+3}_{-2})_{Fp4s} (2)_{\rm FV} (3)_{f_\pi\; {\rm PDG}}\ {\rm fm}\\
      w_0  &=& 0.1732 (4)_{stat} (8)_{extrap} ({}^{+3}_{-2})_{Fp4s} (2)_{\rm FV} (3)_{f_\pi\; {\rm PDG}}\ {\rm fm}\ .
\end{eqnarray*}
Our results for $\sqrt{t_0}$ and $w_0$ differ by 1.7 and 1.1 (joint) sigma, respectively, from BMW's results, which are $\sqrt{t_0}=0.1465(21)(13)$fm and $w_0=0.1755(18)(04)$fm \cite{bmw}. HPQCD \cite{hpqcd} used a subset of the HISQ ensembles included in our computation, but performed an independent gradient flow analysis using the Wilson action in the flow and extrapolating with $f_\pi$ instead of $F_{p4s}$. Our results for $\sqrt{t_0}$ and $w_0$ differ by 0.3 and 1.9, respectively,  of the current sigma from HPQCD's results, which  are $\sqrt{t_0}=0.1420(08)$fm and $w_0=0.1715(09)$fm. 
We do not use joint sigmas in this case because of the correlations between the two determinations.
The difference with HPQCD on $w_0$ can be attributed largely to our inclusion of the $a\approx0.06$ fm ensembles in our analysis, and, to a lesser extent, our inclusion of $\alpha_s a^2$ and $a^2$ terms simultaneously in the continuum extrapolation. The combination of $\alpha_s a^2$ and $a^2$ terms are mainly needed because of the upward ``hook'' in the data for $a<0.09$ fm, as seen in \figref{histextrap}.

\vspace{-2mm}
\acknowledgments
\vspace{-2mm}
This work was supported by the U.S. Department of Energy and the National Science Foundation. Computations for the calculation of gradient flow were carried out with resources provided by the Texas Advanced Computing Center (TACC).  The HISQ gauge configurations were generated with resources provided by the Argonne Leadership Computing Facility, the Blue Waters Project at the National Center for Supercomputing Applications (NCSA), the National Energy Resources Supercomputing Center (NERSC), the National Institute for Computational Sciences (NICS), TACC, the National Center for Atmospheric Research (UCAR), and the USQCD facilities at Fermilab, under grants from the DOE and NSF.


\vspace{-2mm}
\begin{thebibliography}{99}
\vspace{-2mm}
\bibitem{r0} R.~Sommer, Nucl.~Phys.~B {\bf 411} (1994) 839 [{arXiv:9310022}].\\
\bibitem{r1} C. Bernard et al., Phys.~Rev.~D {\bf 62}, (2000) 034503 [{arXiv:hep-lat/0002028}].\\
\bibitem{fp4s} A.~Bazavov {\it et al.}  [MILC Collaboration], Phys.~Rev.~D {\bf 90} (2014) 074509  [{arXiv:1407.3772}].\\
\bibitem{luscher_t0} M.~L\"uscher, J.~High~Energy~Phys. {\bf 1008} (2010) 071 [{arXiv:1006.4518}].\\
\bibitem{bmw} S.~Borsanyi {\it et al.}  [BMW], J.~High~Energy~Phys. {\bf 1209} (2012) 010 [{arXiv:1203.4469}].\\
\bibitem{hisq1} A.~Bazavov {\it et al.}  [MILC Collaboration], Phys.~Rev.~D {\bf 82} (2010) 074501 [{arXiv:1004.0342}].\\
\bibitem{hisq2} A.~Bazavov {\it et al.}  [MILC Collaboration], Phys.~Rev.~D {\bf 87} (2013) 054505 [{arXiv:1212.4768}].\\
\bibitem{smoothing_origin} R.~Narayanan and H.~Neuberger, J.~High~Energy~Phys. {\bf 0603} (2006) 064 [{arXiv:hep-th/0601210}].\\
\bibitem{luscher_origin} M.~L\"uscher, Commun.~Math.~Phys. {\bf 293} (2010) 899 [{arXiv:0907.5491}].\\
\bibitem{chipt} O.~Bar and M.~Golterman, Phys.~Rev.~D {\bf 89} (2014) 034505 [{arXiv:1312.4999}].\\
\bibitem{charm1} W.~Bernreuther and W.~Wetzal, Nucl.~Phys.~B {\bf 197} (1982) 228.\\
\bibitem{charm2} Manohar and Wise, \emph{Heavy Quark Physics} (Cambridge University Press, 2010).\\
\bibitem{prelim} A.~Bazavov {\it et al.}  [MILC Collaboration], \pos{PoS(Lattice 2013)269} [{arXiv:1311.1474}].\\
\bibitem{zeugen} Z.~Fodor {\it et al.}, J.~High~Energy~Phys. {\bf 1409} (2014) 018 [{arXiv:1406.0827}].\\
\bibitem{PDG} J. Beringer et al. [Particle Data Group], Phys.~Rev.~D {\bf 86} (2012) 010001. \\
\bibitem{hpqcd} R.~J.Dowdall {\it et al.} [HPQCD], Phys.~Rev.~D {\bf 88} (2013) 074504 [{arXiv:1303.1670}].\\
\end{thebibliography}
\end{document}